\documentclass[twocolumn,preprintnumbers,amsmath,amssymb,showkeys,aps]{revtex4}
\usepackage{txfonts}
\usepackage{amsfonts}
\usepackage{amsmath}
\usepackage{graphicx}
\usepackage{dcolumn}
\usepackage{bm}
\usepackage{overpic}
\usepackage[dvips]{color}

\begin{document}
\title{Percolation in Interdependent and
Interconnected Networks: \\Abrupt Change from Second to First Order Transition}

\author{Yanqing Hu$^{1,2}$, Baruch Ksherim$^1$, Reuven Cohen$^3$, Shlomo Havlin$^1$}
 \affiliation{1. Department of Physics, Bar-Ilan University, Ramat-Gan 52900, Israel
\\2. Department of Systems Science, School of Management and Center for Complexity
 Research, Beijing Normal University, Beijing 100875, China\\
 3. Department of Mathematics, Bar-Ilan University, Ramat-Gan 52900, Israel}

\date{\today}

\begin{abstract}
Robustness of two coupled networks system has been studied only for
dependency coupling (S. Buldyrev et. al., Nature, 2010) and only for
connectivity coupling (E. A. Leicht and R. M. D'Souza,
arxiv:09070894). Here we study, using a percolation approach, a more
realistic coupled networks system where both interdependent and
interconnected links exist. We find a rich and unusual phase
transition phenomena including hybrid transition of mixed first and
second order i.e., discontinuities like a first order transition of
the giant component followed by a continuous decrease to zero like a
second order transition. Moreover, we find unusual discontinuous
changes from second order to first order transition as a function of
the dependency coupling between the two networks.
\end{abstract}



\maketitle

During the last decade complex networks have been studied
intensively, where most of the research was devoted to analyzing the
structure and functionality isolated systems modeled as single
non-interacting networks
\cite{barasci,ws,bararev,Yanqing2011,DPoCN,NRaFPoRG,PastorXX,mendes,cohena,Newman-book,song}.
However, most real networks are not isolated, as they either
complement other networks (``interconnected networks''), must
consume resources supplied by other networks ("interdependent
networks") or both
\cite{Rosato2008,Nils2007,Rinaldi2001,Peerenboom2001,laprie}. Thus,
real networks continuously interact one with each other, composing
large complex systems, and with the enhanced development of
technology, the coupling between many networks becomes more and more
significant.

Two different types of coupled networks models have been studied.
Buldyrev et. al. \cite{SbRpGpHesSh} investigated the robustness of
coupled systems with only interdependence links. In these systems,
when a node of one network fails, its dependent counterpart node in
the other network also fails. They found that this interdependence
makes the system significantly more \textit{vulnerable}
\cite{SbRpGpHesSh,Parshani}. In the same time, Leicht and D'Souza
\cite{Leicht} studied the case where only connectivity links couple
the networks, i.e., ``interconnected networks'', and found that the
interconnected links make the system significantly more
\textit{robust}. However, real coupled networks often contain both
types of links, interdependent as well as interconnected links. For
example, the airport and the railway networks in Europe are two
coupled networks composing a transportation system. In order to
arrive to an airport, one usually uses the railway. Also, people
arriving to the country by airport usually use the railway. In this
system, if the airport is disabled by some strike or accident, the
passengers can still use the nearby railway station and travel to
their destination or to another airport by train, so the two
networks are coupled by connectivity links. On the other hand, if
the railway network is disabled, the airport traffic is damaged, and
if the airport is disabled, the railway traffic is damaged, so both
networks are coupled by dependency links as well. The important
characteristics of such systems, is that a failure of nodes in one
network carries implications not only for this network, but also on
the function of other dependent networks. In this way it is possible
to have cascading failures between the coupled networks, that may
lead to a catastrophic collapse of the whole system. Nevertheless,
small clusters disconnected from the giant component in one network
can still function through interconnected links connecting them to
the giant component of other network. Thus, the inter-connectivity
links \textit{increase} the robustness of the system, while the
inter-dependency links \textit{decrease} its robustness. Here we
study the competition of the two types of inter-links on robustness
using a percolation approach, and find unusual types of phase
transitions.

\begin{figure}
\includegraphics[width=5cm]{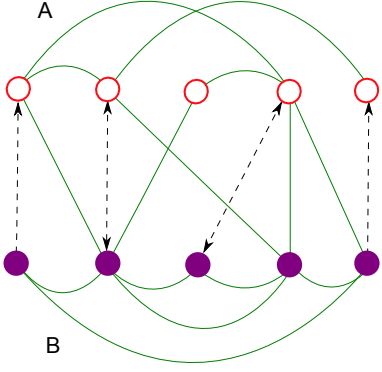}
\caption{(Color online) Two types of inter-links where the
dependency links (dashed arrows) are not necessarily bidirectional.
The nodes of A and B are randomly connected with {\em
connectivity-links} (full line). The functionality of some of the A
nodes (red open circle) \emph{depend} on B-nodes (purple solid
circle) and vice versa.}\label{dependencelinks}
\end{figure}

Let us consider a system of two networks, $A$ and $B$, which are
coupled by both dependency and connectivity links. The two networks
are partially coupled by dependency links, so that a fraction $q_A$
of $A$-nodes depends on nodes in network $B$, and a fraction $q_B$
of $B$-nodes depends on the nodes in network $A$, with the following
two exceptions: a node from one network depends on no more than one
node from the other network, and assuming that node $A_i$ depends on
node $B_j$, then if $B_j$ depends on some $A_h$, then $h=i$ (see
Fig.~\ref{dependencelinks}). In addition, the connectivity links
within each network and between the networks (see Fig.
\ref{dependencelinks}) can be described by a set of degree
distributions $\{\rho^{A}_{k_A,k_{AB}},\rho^{B}_{k_{B},k_{BA}}\}$,
where $\rho^{A}_{k_{A},k_{AB}}$ ($\rho^{B}_{k_{B},k_{BA}}$) denotes
the probability of an $A$-node ($B$-node) to have $k_A$ ($k_{B}$)
links to other $A$-nodes ($B$-node) and $k_{AB}$ ($k_{BA}$) links
towards $B$-nodes ($A$-nodes). In this manner we get a two
dimensional generating function describing all the connectivity
links \cite{Leicht}, $\mathcal
G^{A}_{0}(x_A,x_B)=\sum\limits_{k_A,k_{AB}}\rho^{A}_{k_{A},k_{AB}}x_{A}^{k_A}x_{B}^{k_{AB}}$,
and $\mathcal
G^{B}_{0}(x_A,x_B)=\sum\limits_{k_B,k_{BA}}\rho^{B}_{{k_B},k_{BA}}x_{A}^{k_{BA}}x_{B}^{k_B}$.

The cascading process is initiated by randomly removing a fraction
$1-p$ of the $A$-nodes and all their connectivity links. Because of
the interdependence between the networks, the nodes in network $B$
that depend on the removed $A$-nodes are also removed along with
their connectivity links. As nodes and links are removed, each
network breaks up into connected components (clusters). We assume
that when the network is fragmented, the nodes belonging to the
largest component (giant component) connecting a finite fraction of
the network are still functional, while nodes that are parts of the
remaining smaller clusters become dysfunctional, unless there exist
a path of connectivity-links connecting these small clusters to the
largest component of the other network. Since the networks have
different topologies, the removal of nodes and related dependency
links, is not symmetric in both networks, so that, a cascading
process occurs, until the system either becomes fragmented or
stabilizes with a giant component.

Let $g_A(\varphi,\phi)$ and $g_B(\varphi,\phi)$ be the fraction of
A-nodes and B-nodes in the giant components after the percolation
process initiated by removing a fraction of $1-\varphi$ and $1-\phi$
of networks A and B respectively \cite{Newman-book}. The functions
$g_A(\varphi,\phi)$ and $g_B(\varphi,\phi)$ depend only on $\mathcal
G^{A}_{0}(x_A,x_B)$ and $\mathcal G^{B}_{0}(x_A,x_B)$ (For details
see SI) and the cascading process can be described by the following
set of equations,
\begin{align}
&\varphi_{1}=p, \quad \phi_{1}=1, &P^A_{1}=\varphi_{1}g_A(\varphi_{1},\phi_{1}),
\\\nonumber &\phi_{2}=1-q_B\Bigl(1-pg_A(\varphi_{1},\phi_{1})\Bigr),
&P^B_{2}=\phi_{2}g_B(\varphi_{1},\phi_{2}),
\\\nonumber&\varphi_{2}=p\Bigl(1-q_A\left(1-g_B(\varphi_{1},\phi_{2})\right)\Bigr), &P^A_{2}=\varphi_{2}g_A(\varphi_{2},\phi_{2}),
\\\nonumber &\phi_{3}=1-q_B\Bigl(1-pg_A(\varphi_{2},\phi_{2})\Bigr),
&P^B_{3}=\phi_{3}g_B(\varphi_{2},\phi_{3}),
\end{align}
where, $\phi_{i}, \varphi_{i}$ are the remaining fraction of nodes
at stage $i$ of the cascade of failures and $P^A_{i}$, $P^B_{i}$ are
the corresponding giant components of networks $A$ and $B$,
respectively. Generally, the $n^{\underline{th}}$ step is given by
the equations,
\begin{align}
&\varphi_{n}=p\Bigl(1-q_A\left(1-g_B(\varphi_{n-1},\phi_{n})\right)\Bigr),
\\\nonumber &\phi_{n}=1-q_B\Bigl(1-pg_A(\varphi_{n-1},\phi_{n-1})\Bigr),
\\\nonumber &P^A_{n}=\varphi_{n}g_A(\varphi_{n},\phi_{n}),\quad
P^B_{n}=\phi_{n}g_B(\varphi_{n-1},\phi_{n}).
\end{align}
By introducing two new notations
\begin{align}
u_A=g_{A}(\phi_{\infty},\varphi_{\infty}),
\quad u_B=g_{B}(\phi_{\infty},\varphi_{\infty}),
\end{align}
we can write the equations at the end of the cascading process,
\begin{align}\label{eq:XYsystem}
\phi_{\infty}=p\Bigl(1-q_A(1-u_B)\Bigr),\quad
 \varphi_{\infty}=1-q_B(1-pu_A),
\end{align}
and the giant components are,
\begin{align}\label{Pinfab}
&P_{\infty}^{A}=u_A\phi_{\infty}=u_Ap\Bigl(1-q_A(1-u_B)\Bigr),
\\\nonumber &P_{\infty}^{B}=u_B\varphi_{\infty}=u_B\Bigl(1-q_B(1-pu_A)\Bigr).
\end{align}

In the case where all degree distributions of intra- and inter-links
are \textit{Poisson} distributed, the functions obtain a simple
form. Assume $\overline{k}_A$ and $\overline{k}_B$ are the average
intra-links degrees in networks $A$ and $B$, and
$\overline{k}_{AB}$, $\overline{k}_{BA}$ are the average inter-links
degrees between $A$ and $B$ (allowing the case
$\overline{k}_{AB}\neq \overline{k}_{BA}$, since the two networks
may be of different sizes), we obtain,
\begin{align}\label{eq:test}
&u_A=1-e^{-\overline{k}_Apu_A\bigl(1-q_A(1-u_B)\bigr)-
\overline{k}_{AB}u_B\bigl(1-q_B(1-pu_A)\bigr)},
\\\nonumber &u_B=1-e^{-\overline{k}_{BA}pu_A\bigl(1-q_A(1-u_B)\bigr)-
\overline{k}_Bu_B\bigl(1-q_B(1-pu_A)\bigr)}.
\end{align}

\begin{figure}
\center
\includegraphics[width=7cm]{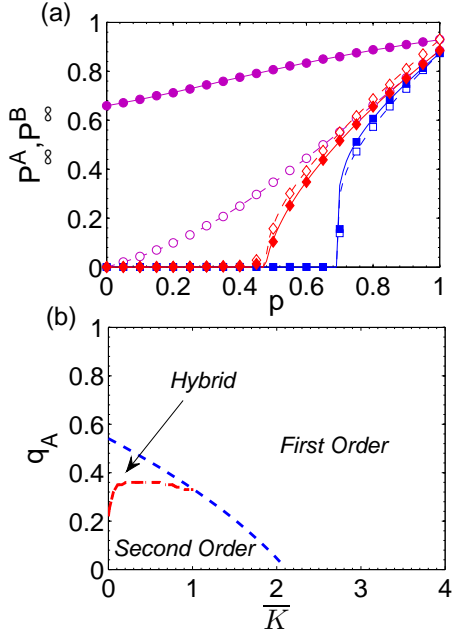}
\caption{(Color online) \textbf{a.} Giant components
$P^{A}_{\infty}$ and $P^{B}_{\infty}$ vs. fraction of remaining
nodes, $p$, for $N=10000$, $\overline{k}=2$ and $\overline{K}=1$.
Networks $A$ (open symbols) and $B$ (full symbols) for different
$(q_A,q_B)$ pairs: $(0.8,0.1)$ ($\circ$); $(0.8,0.8)$ ($\diamond$);
$(0.1,0.1)$ ($\square$). The symbols represent simulations and the
lines the theory. We see three types of behaviors: no phase
transition ($\circ$), second order phase transition ($\diamond$) and
first order phase transition ($\square$). \textbf{b.} Phase diagram
showing the first order, second order and hybrid phase transition
regimes and the boundaries, for $q_B=1, \overline{k}=3$. In the
second order transition regime, between the two dashed curve (red
and blue) is the hybrid phase transition regime (details in
Fig.~\ref{simu_And_anlysis2}\textbf{c} and in the SI). Since the
hybrid transition is continuous in the neighborhood of $p_c$, and
jump occurs well above $p_c$ we classify a hybrid phase transition
as a second order phase transition.} \label{simu_And_anlysis1}
\end{figure}

Generally, for fixed parameters $\overline{k}_A, \overline{k}_B,
\overline{k}_{AB}, \overline{k}_{BA}, q_A, q_B$ and $p$, it is often
impossible to achieve an explicit formula for the giant components
$P_{\infty}^A$ and $P_{\infty}^B$. However, one can still solve
Eqs.~(\ref{eq:test}) graphically and substitute the numerical
solution to Eqs.~(\ref{Pinfab}). For example, we study the case
where $\overline{k}_A=\overline{k}_B\equiv\overline{k}$ and
$\overline{k}_{AB}=\overline{k}_{AB}\equiv \overline{K}$.
Fig.~\ref{simu_And_anlysis1}\textbf{a} compares the numerical with
the simulation results for $P^A_{\infty}$ and $P^B_{\infty}$ as a
function of $p$, showing that the analytical results of
Eqs.~(\ref{Pinfab}) and (\ref{eq:test}) are in excellent agreement
with the simulations.

Next we are interested in the properties of the phase transition
under random attack, so first we determine the conditions when
transition does not occur. This is the case when even all nodes of
network A are removed $(p=0)$, for a given $q_B<1$, there still
exists a giant component in network B (see circles in
Fig.~\ref{simu_And_anlysis1}\textbf{a}) and no phase transition
occurs. For Poisson degree distributions, if after the removal of
all B-nodes that depend on the attacked A-nodes, the new average
intra-link degree in network B is less than one, i.e.,
\begin{equation}
\overline{k}_B(1-q_B)<1, \label{notransition}
\end{equation}
a phase transition occurs. Therefore, the following analysis is
based on condition~\eqref{notransition}. In addition, we always set
both dependency strengths, $q_A$ and $q_B$, to be larger than zero.

\begin{widetext}
\center
\begin{figure}
\includegraphics[width=1\columnwidth]{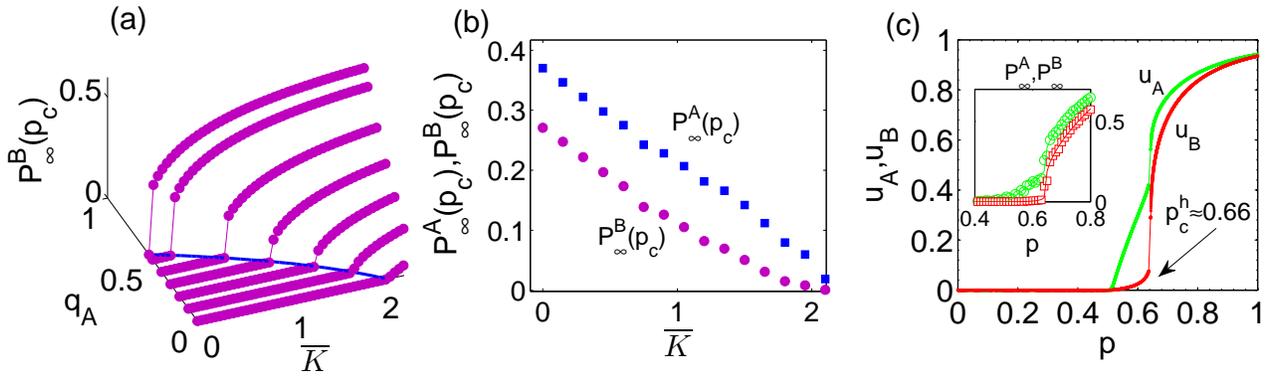}
\caption{\textbf{a.} Size of giant components vs. dependency and
connectivity links strength, for $q_B=1$ and $\overline{k}=3$. The
giant components size at $p_c$ changes from zero to a finite value
while changing $q_A$ and $\overline{K}$. When $q_A$ and
$\overline{K}$ are at the boundary of different phase transitions,
the jump occurs. \textbf{b.} The values of $P^A_{\infty}(p_c)$
($\circ$), $P^B_{\infty}(p_c)$ ($\square$) along the boundary for
$q_B=1$ and $\overline{k}=3$. \textbf{c.} Hybrid phase transition,
for $q_B=1, q_A=0.35$, $\overline{k}=3$ and $\overline{K}=0.1$.
According to Eqs.~(\ref{Pinfab}), $P^A_{\infty}$ and $P^B_{\infty}$
have the same properties as $u_A$ and $u_B$ respectively. At
$p\approx0.66$ the values of $u_A$ and $u_B$ jump, and then for
lower $p$ valve continuously approach zero. In the inset, simulation
and theoretical results are symbols and lines
respectively.}\label{simu_And_anlysis2}
\end{figure}
\end{widetext}
When the phase transition is of second order, i.e., the giant
components at the percolation threshold is zero. Thus, according to
the limit of system (\ref{eq:test}) at $u_A=u_B=0$ we obtain the
second order threshold, for $q_A\neq1$,
\begin{equation}\label{cpc}
p^{II}_c=\frac{1-\overline{k}_B(1-q_{B})}{\Bigl(\overline{k}_A+(\overline{k}_{BA}\overline{k}_{AB}-k_Ak_B)(1-q_B)\Bigr)(1-q_A)}.
\end{equation}
When $q_A=1$ and $0\leq q_B<1$ this threshold becomes
\begin{equation}
\nonumber p^{II}_c=\frac{1}{\overline{k}_B(1-q_B)}>1,
\end{equation}
which together with Eq. (\ref{notransition}) implies that the phase
transition must be of first order  at $p_c^I<1$ that will be
determined later.

Solving the first equation of system (\ref{eq:test}), yields an
explicit formula for $u_B$, so that system (\ref{eq:test}) can be
rewritten as
\begin{align}\label{tangentialsystem}
&u_B=-\frac{\log(1-u_A)+k_Ap(1-q_A)u_A}{k_Apq_Au_A+k_{AB}[1-q_B(1-pu_A)]}\equiv
H_1(u_A),
\\\nonumber &u_B=1-e^{-\overline{k}_{BA}u_Ap\bigl(1-q_A(1-u_B)\bigr)-
\overline{k}_Bu_B\bigl(1-q_B(1-u_Ap)\bigr)}\equiv H_2(u_A).
\end{align}
and the intersection of the two curves (maximum solution of $u_A,
u_B$) is the solution of the system. When the phase transition is
first order and $p=p_c^{I}$, the curves of
Eqs.~\eqref{tangentialsystem} are tangentially touching at the
solution point, where,
\begin{equation}
\Bigl(\frac{dH_1}{du_A}=\frac{dH_2}{du_A}\Bigr)\bigg|_{p=p_c^{I}}.
\label{p1}
\end{equation}
Obviously, $u_A,~u_B$ and $p$ can be treated as variables of
Eqs.~(\ref{tangentialsystem}) and (\ref{p1}). Solving these
equations, the minimal solution of $p$ and the corresponding maximum
$u_A,~u_B$ of the minimal $p$ is the
solution of the system at criticality.

When networks A and B are fully dependent, i.e., $q_A=q_B=1$, system
(\ref{eq:test}) yields a simple form
\begin{align}
\nonumber u_A=1-\exp\Bigl\{-pu_Au_B\Bigl(\overline{k}_A+\overline{k}_{AB}\Bigr)\Bigr\},
\\\nonumber u_B=1-\exp\Bigl\{-pu_Au_B\Bigl(\overline{k}_B+\overline{k}_{BA}\Bigr)\Bigr\}.
\end{align}
The size of the mutual giant component, $P_{\infty}$, is thus given
by,
\begin{equation}
P_{\infty}=P_{\infty}^{A}=P_{\infty}^{B}=p\Bigl(1-e^{-P_{\infty}(\overline{k}_A+\overline{k}_{AB})}\Bigr)\Bigl(1-e^{-P_{\infty}(\overline{k}_B+\overline{k}_{BA})}\Bigr),
\end{equation}
which is similar to the solution of fully interdependent system
\cite{SbRpGpHesSh}, where the only difference is that the degrees of
networks A and B are now replaced by
$\overline{k}_A+\overline{k}_{AB}$ and
$\overline{k}_B+\overline{k}_{BA}$, respectively. Thus,
interestingly, in a fully interdependent coupled networks adding
connectivity inter-links has the same effect as increasing the
intra-degree of the corresponding networks and therefore, in this
case, the phase transition must be of first order. From
Eqs.~\eqref{tangentialsystem} and \eqref{p1}, one can get the
threshold,
\begin{equation}
p_c^{I}=\frac{1}{{k_A(1-u_A)\Bigl[-1+(1-u_A)^{\alpha}-u_A\alpha(1-u_A)^{\alpha-1}\Bigr]}},
\end{equation}
where,
$\alpha\equiv(\overline{k}_B+\overline{k}_{BA})/(\overline{k}_A+\overline{k}_{AB})$,
and $u_A$ satisfies the equation,
\begin{equation}
u_A=1-\exp\Bigl\{\frac{u_A[1-(1-u_A)^{\alpha}]}{(1-u_A)[-1+(1-u_A)^{\alpha}-u_A\alpha(1-u_A)^{\alpha-1}]}\Bigr\}.
\end{equation}
For fully interdependent system, both networks are of the same size
and therefore $\overline{k}_{AB}=\overline{k}_{BA}$.

By substituting $p^{II}_{c}$ from Eq. (\ref{cpc}) into
Eqs.~\eqref{tangentialsystem} and \eqref{p1} and evaluating both
$u_A$ and $u_B$ we can derive and draw in the phase diagram, the
boundary between the first and second order transitions (see dashed
line in Fig.~\ref{simu_And_anlysis1}\textbf{b}). The most
interesting phenomenon, which to the best of our knowledge, has not
been observed before, is that when the phase transition changes from
first to second, there are discontinuities (abrupt jumps) of
$P^{A}_{\infty}(p_c), P^{B}_{\infty}(p_c)$ in the phase transition
boundary (see Fig.~\ref{simu_And_anlysis2}\textbf{a}). The values of
the jumps along the boundary are shown in
Fig.~\ref{simu_And_anlysis2}\textbf{b} (details in Fig. 1 of SI).
This phenomenon contrasts most systems possessing both first and
second order transitions. In physical systems usually, the first
order jump in the order parameter, and related properties, such as
the specific heat, present a continuous change along the transition
line when the
system changes from first to second order. 

In addition to the existence of jumps in $P^{A}_{\infty}(p_c),
P^{B}_{\infty}(p_c)$ at the boundary between the first and second
order phase transitions, we find another unusual phenomenon. When
one network strongly depends on the other, there exist hybrid phase
transitions. By hybrid phase transition we mean that when increasing
the attack strength, $1-p$, the size of the giant component jumps at
$p_c^h$ from a large value to a small value, and then continuously
decreases to zero. A similar behavior has been found in bootstrap
percolation \cite{hybrid}. Since the second order transition is
characterized by a giant component which is continuous in the
neighborhood of $p_c$, we regard, the hybrid phase transition regime
as a second order phase transition regime (see
Fig.~\ref{simu_And_anlysis1}\textbf{b}). For the hybrid phase
transition, there exists a threshold $p_c^{h}$ at which the jump
occurs (see Fig.~\ref{simu_And_anlysis2}\textbf{c}). For $p$ just
below $p_c^{h}$, the solution of Eqs.~(\ref{tangentialsystem}) for
$u_A,~u_B$, will jump to lower values (For more details see Chap. 3
in SI). After the jump, when $p$ is further decreased, $u_A$ and
$u_B$ approach to zero continuously which implies that the giant
components sizes change to zero continuously. For example, for the
parameters $q_A=0.35, q_B=1,\overline{k}=3$ and $\overline{K}=0.1$,
we obtain $p_c=0.556$ and $p_c^{h}=0.66$. When $p$ is just below
0.66, the giant components drops to smaller positive values like in
a first order phase transition. After this discontinuous drop, the
giant component's size continuously decreases to zero when
decreasing $p$ from 0.66 to 0.556 like a second order phase
transition (see Fig.~\ref{simu_And_anlysis2}\textbf{c}).

In summary, we studied the cascade of failures in coupled networks,
when both interdependent and interconnected links exist, using a
percolation approach. Although our detailed analysis is for ER
networks, the theory can be applied to any network systems topology.
We find that the existence of inter-connectivity links between
interdependent networks, introduces rich and intriguing phenomena
through the process of cascading failures. Increasing the strength
of interconnecting links can change the transition behavior
significantly and often brings up some counterintuitive phenomenon,
such as changing the transition from second order to first order (as
seen in Fig.~\ref{simu_And_anlysis1}\textbf{b}). We also find an
unusual abrupt jump in the boundary between first and second order
phase transitions at the critical point, which, to the best of our
knowledge, has not been observed earlier in physical systems.
Moreover, when one of the networks strongly depends on the other
network, unusual hybrid phase transitions are observed.

We thank Amir Bashan for helpful discussions. This work is partially
supported by ONR, DFG, DTRA, EU project Epiwork and the Israel
Science Foundation for financial support. Y. Hu is supported by NSFC
under Grant No. 60974084, 60534080.

\textbf{Supplementary Information}

\section{How to get $g_A(\varphi,\phi)$ and $g_B(\varphi,\phi)$}

We model the percolation process using the branching process
approach. Let $\mathcal
G^{A}_{0}(x_A,x_B)=\sum\limits_{k_A,k_{AB}}\rho^{A}_{k_{A},k_{AB}}x_{A}^{k_A}x_{B}^{k_{AB}}$,
and $\mathcal
G^{B}_{0}(x_A,x_B)=\sum\limits_{k_B,k_{BA}}\rho^{B}_{{k_B},k_{BA}}x_{A}^{k_{BA}}x_{B}^{k_B}$,
be the degree distributions' generating functions. The probability
of following a randomly chosen $AB$-link connecting an $A$-node of
degree $k_{A}$ to a $B$-node with excess $k_{AB}$ degree (i.e.,
having total $A$ to B degree of $k_{AB}+1$) is proportional to
$(k_{AB}+1)\rho^{A}_{k_{A},k_{AB}}$, and the generating function for
this distribution is,
\begin{align}
\mathcal
G^{AB}_{1}(x_A,x_B)=\sum_{k_A,k_{AB}}\frac{(k_{AB}+1)\rho^{A}_{k_{A},k_{AB}+1}}
{\sum\limits_{k^{\prime}_{A},k^{\prime}_{AB}}k^{\prime}_{AB}
\thinspace\rho^{A}_{k^{\prime}_{A},k^{\prime}_{AB}}}\cdot
x_{A}^{k_A}x_{B}^{k_{AB}}\;.
\end{align}
Analogously, we construct the other three excess generating
functions $\mathcal G^{AA}_{1}(x_A,x_B),\mathcal
G^{BA}_{1}(x_A,x_B)$ and $ \mathcal G^{BB}_{1}(x_A,x_B)$.

After removing a fraction $1-\varphi$ of nodes in network $A$, and a
fraction $1-\phi$ of nodes in network $B$, we can set new arguments
to the generating functions, so that, $x_A$ and $x_B$ will be
replaced by $1-\varphi(1-x_A)$ and $1-\phi(1-x_B)$, respectively
\cite{NewmanSpread,ShaoEPL,ShaoPRE}. Suppose
$g_A(\varphi,\phi),~g_B(\varphi,\phi)$ are the fractions of
$A$-nodes and $B$-nodes in the giant components after removal of
$1-\varphi$ and $1-\phi$ fractions of networks $A$ and $B$,
respectively. Then we have,
\begin{align}\label{eq:gAgB}
&g_A(\varphi,\phi)=1-\mathcal
G_{0}^{A}\Bigl(1-\varphi(1-f_{A}),1-\phi(1-f_{BA})\Bigr),
\\\nonumber &g_B(\varphi,\phi)=1-\mathcal G_{0}^{B}\Bigl(1-\varphi(1-f_{AB}),1-\phi(1-f_{B})\Bigr),
\end{align}
where,
\begin{align}\label{f}
&f_{A}=\mathcal
G_{1}^{AA}\Bigl(1-\varphi(1-f_{A}),1-\phi(1-f_{BA})\Bigr),
\\\nonumber &f_{AB}=\mathcal G_{1}^{AB}\Bigl(1-\varphi(1-f_{A}),1-\phi(1-f_{BA})\Bigr),
\\\nonumber &f_{BA}=\mathcal G_{1}^{BA}\Bigl(1-\varphi(1-f_{BA}),1-\phi(1-f_{B})\Bigr),
\\\nonumber &f_{B}=\mathcal G_{1}^{BB}\Bigl(1-\varphi(1-f_{BA}),1-\varphi(1-f_{B})\Bigr).
\end{align}

When all of the degree distributions of inter and intra networks $A$
and $B$ are Poisson distribution, all of the functions can be more
simple. Assume $\overline{k}_A$ and $\overline{k}_B$ are the average
intra-links degrees in networks $A$ and $B$ and $\overline{k}_{AB}$,
$\overline{k}_{BA}$ are the average inter-links degrees between $A$
and $B$ (allowing the case $\overline{k}_{AB}\neq
\overline{k}_{BA}$, since the network sizes of $A$ and $B$ can be
different), then we have
$G_{0}^{AA}(x_A)=e^{\overline{k}_A(x_A-1)}$,
$G_{0}^{AB}(x_B)=e^{\overline{k}_{B}(x_B-1)}$,
$G_{0}^{BA}(x_A)=e^{\overline{k}_{BA}(x_A-1)}$,
$G_{0}^{BB}(x_B)=e^{\overline{k}_{B}(x_B-1)}$ and
\begin{align}
\\\nonumber\mathcal G^{AA}_{1}(x_A,x_B)=\mathcal
G^{AB}_{1}(x_A,x_B)=G^{A}_{0}(x_A,x_B)=G_{0}^{AA}(x_A)G_{0}^{AB}(x_B)
\\\nonumber \mathcal G^{BB}_{1}(x_A,x_B)=\mathcal G^{BA}_{1}(x_A,x_B)=G^{B}_{0}(x_A,x_B)=G_{0}^{BA}(x_A)G_{0}^{BB}(x_B)
\end{align}
Submitting above equations to to systems.~(\ref{eq:gAgB}) and
(\ref{f}), we get
\begin{align}\label{eq:g_A_g_B}
&g_{A}(\varphi,\phi)=1-\exp\Bigl\{-\overline{k}_Axg_{A}(\varphi,\phi)-\overline{k}_{AB}yg_{B}(\varphi,\phi)\Bigr\},
\\\nonumber &g_{B}(\varphi,\phi)=1-\exp\Bigl\{-\overline{k}_{BA}xg_{A}(\varphi,\phi)-\overline{k}_Byg_{B}(\varphi,\phi)\Bigr\}.
\end{align}

\section{Abrupt jump on the boundary}
We rewrite the main system here
\begin{align}\label{tangentialsystem}
&u_B=-\frac{\log(1-u_A)+\overline{k}_Ap(1-q_A)u_A}{\overline{k}_Apq_Au_A+\overline{k}_{AB}[1-q_B(1-pu_A)]}\equiv
H_1(u_A),
\\\nonumber &u_B=1-e^{-\overline{k}_{BA}u_Ap\bigl(1-q_A(1-u_B)\bigr)-
\overline{k}_Bu_B\bigl(1-q_B(1-u_Ap)\bigr)}\equiv H_2(u_A).
\\\nonumber &\Bigl(\frac{dH_1}{du_A}=\frac{dH_2}{du_A}\Bigr)\bigg|_{p=p_c^{I}}.
\end{align}
On the boundary between first and second order phase transition,
$p_c^{I}=p_c^{II}$. Substituting $p_c^{I}$ with $p^{II}_{c}$ in
system~\eqref{tangentialsystem} and evaluating both $u_A$ and $u_B$
we can obtain the boundary between the first and second order
transitions. When we reduce the three equations to one equation,
$u_A,~u_B$ should always be the maximum non-negative solution in
[0,1]. When system.~\eqref{tangentialsystem} has more than one
solution, we always choose the minimal non-negative value,
$p_c^{min}$ and the corresponding maximum value solution
$u_A^{max},~u_B^{max}$ as the solution at the threshold. In some
regime of the boundary, $u_A^{max}>0$ and $u_B^{max}>0$, and of
course $p_c^{min},u_A=0,~u_B=0$ also is the system solution. It
means that there exist two intersections and both of them satisfy
the tangential condition (as shown in Fig. \ref{tangentialsystem})
on the boundary. This implies that when the order of the phase
transition changes from first to second,
$P^{A}_{\infty}\Bigl(p_c\Bigr), P^{B}_{\infty}\Bigl(p_c\Bigr)$ are
discontinuous.

\section{Hybrid Phase transition}
The minimum solution of $p^{min}$ in [0,1] of
system~\eqref{tangentialsystem} is the $p_c$. Besides $p^{min}$ if
system~\eqref{tangentialsystem} has another solution
$p_c^{h}\in(0,1)$ and corresponding solution $u_A^{h},~u_B^{h}$, we
can find the hybrid phase transition. $(p^{h},~u_A^{h},~u_B^{h})$
means that when $p$ is little less than $p_c^{h}$, the solution
$u_A,~u_B$ of the first two equations of
system~(\ref{tangentialsystem}) will jump to small values. After the
jump, when we continue to decrease $p$ to $p_c=p^{min}$, $u_A,~u_B$
will move to 0 continually (as shown in Fig. \ref{CJ}).

\begin{figure}
\includegraphics[width=0.7\columnwidth]{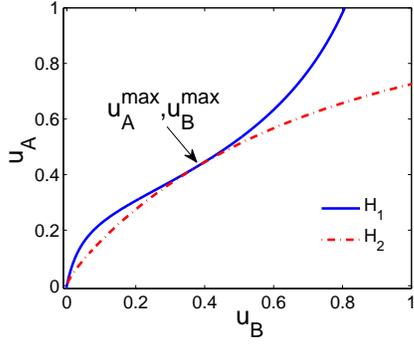}
\caption{Abrupt jump on the boundary, here $q_A=0.394, q_B=0.8,
\overline{k}=3, \overline{K}=0.2$. $p_c^{I}=p_c^{II}=0.5464$ which
is the threshold of the system. Although, both intersections (one of
which is at the origin) satisfy the tangential condition, the
$u_A^{max},u_B^{max}$ is the physical solution and the transition is
of the first order.}\label{0OC}
\end{figure}

\begin{figure}
\includegraphics[width=.8\columnwidth]{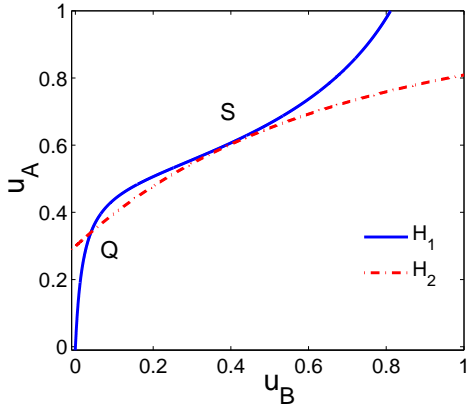}
\caption{Hybrid transition analysis, for $q_B=1,
q_A=0.35,\overline{k}=3$ and $\overline{K}=0.1$, here
$p_c\approx0.556$. $p^h\approx0.66$. The maximum intersection S
satisfies tangential condition. When continuously decreasing $p$,
the solution of the system jumps from the maximum intersection S to
the minimum intersection Q and then continuously decrease to zero.
}\label{CJ}
\end{figure}

\end{document}